# Probing ultrafast spin-relaxation and precession dynamics in a cuprate Mott insulator with 7-fs optical pulses

T. Miyamoto[1], Y. Matsui[1], T. Terashige[2], T. Morimoto[1], N. Sono[1], H. Yada[1], S. Ishihara[3], Y. Watanabe[4], S. Adachi[4], T. Ito[5], K. Oka[5], A. Sawa[5], H. Okamoto[1,2*]

[1]Department of Advanced Materials Science, University of Tokyo, Chiba 277-8561, Japan.

[2]AIST-UTokyo Advanced Operando-Measurement Technology Open Innovation Laboratory (OPERANDO-OIL), National Institute of Advanced Industrial Science and Technology (AIST), Chiba 277-8568, Japan

[3]Department of Physics, Tohoku University, Sendai 980-8578, Japan.

[4]Department of Chemistry, Kyoto University, Kitashirakawa Oiwake-cho, Sakyo-ku Kyoto, 606-8502, Japan.

[5]National Institute of Advanced Industrial Science and Technology, Tsukuba, Ibaraki 305-8565, Japan

*Correspondence to H. Okamoto (okamotoh@k.u-tokyo.ac.jp)

Abstract

A charge excitation in a two-dimensional Mott insulator is strongly coupled with the surrounding spins, which is observed as magnetic-polaron formations of doped carriers and a magnon sideband in the Mott-gap transition spectrum. However, the dynamics related to the spin sector are difficult to measure. Here, we show that pump-probe reflection spectroscopy with 7-fs laser pulses can detect the optically induced spin dynamics in $Nd_2CuO_4$, a typical cuprate Mott insulator. The bleaching signal at the Mott-gap transition is enhanced at ~18 fs, which corresponds to the spin-relaxation time in magnetic-polaron formations and is characterized by the exchange interaction. More importantly, ultrafast coherent oscillations appear in the time evolutions of the reflectivity changes, and their frequencies (1400–2700 $cm^{-1}$) are equal to the probe energy measured from the Mott-gap transition peak. These oscillations originate from interferences between charge excitations with two magnons and provide direct evidence for charge-spin coupling.

Carrier doping induces Mott-insulator to metal transitions in various transition-metal oxides[1]. Typical examples are layered cuprates such as $La_2CuO_4$ and $Nd_2CuO_4$[2-6]. In such cuprates, charge-spin coupling affects a charge carrier strongly, which moves on the antiferromagnetic spin background[7-10]. Theoretical studies predict that an isolated carrier forms a kind of magnetic polaron via spin-orientation changes[11-14] and that the Mott-gap transition peak in the optical conductivity ($\sigma$) spectrum is accompanied by a two-magnon sideband via charge-spin coupling[15]. However, information about spin dynamics associated with magnetic-polaron formation and the two-magnon excitation has yet to be observed. Femtosecond pump-probe (PP) spectroscopy may address these unsolved issues since it can detect the time characteristics of photoexcited states[16-21], helping to derive spin-dynamics information.

The studied material is $Nd_2CuO_4$. The Cu-O plane of $Nd_2CuO_4$ is schematically shown in Fig. 1a. The copper $3d_{x^2-y^2}$ and oxygen $2p_{x,y}$ orbitals form the two-dimensional electronic state. In the copper $3d$ band, the Mott-Hubbard gap is opened due to the large on-site Coulomb repulsion. The occupied oxygen $2p$ band is located between the copper $3d$ upper- and lower-Hubbard bands. The lowest electronic excitation is the charge-transfer (CT) transition from the oxygen $2p$ band to the copper $3d$ upper-Hubbard band. Figure 1b shows the reflectivity ($R$) and $\sigma$ spectra of $Nd_2CuO_4$ in which the CT transition is observed as the peak structure. Hereafter, the CT transition is simply called the Mott-gap transition. The strong $pd$ hybridization leads to a large antiferromagnetic exchange interaction $J\sim 0.155$ eV between two neighboring copper spins[22].

When electron carriers are introduced into $CuO_2$ planes by substituting Nd with Ce in $Nd_{2-x}Ce_xCuO_4$, the spectral weight of the Mott-gap transition is transferred to the mid-gap absorption[2-4] originating from magnetic polarons, and the Drude component is enhanced for $x > 0.1$. However, the Drude weight is small compared to the mid-gap absorption even in optimally doped samples exhibiting superconductivity[3,4]. In $Nd_2CuO_4$, the photocarrier dynamics have been investigated by PP absorption spectroscopy with time resolutions of 40–200 fs[16,17]; a photoinduced Mott-insulator to metal transition was demonstrated, but the spin-relaxation processes of photocarriers were not detected due to the insufficient time resolution. Note that the time scale of the spin dynamics evaluated from $J$~0.155 eV is ~30 fs.

In this study, we perform PP reflection spectroscopy on $Nd_2CuO_4$ single crystals with ultrashort pulses width the temporal width of 7 fs, which is illustrated in the inset of Fig. 1b. Those pulses are generated from a handmade non-collinear optical parametric amplifier. The cross-correlation profile of the pump and probe pulses is shown in Fig. 1c. The temporal width of this profile, which is about 10 fs, corresponds to the time resolution in our system. This pulse is divided into two pulses, which are used as pump and probe pulses. The spectrum of the pulse is shown by the colored area in Fig. 1b, which is located within the Mott-gap-transition band. By detecting the transient decrease in the reflectivity of the probe pulse, that is, the bleaching signal, we can investigate the photocarrier dynamics.

**Results**

**Excitation-photon-density dependence on the time evolutions of the reflectivity changes.**

Figure 2a shows the time evolutions of the reflectivity changes $\Delta R(t)/R$ of the probe pulse for the excitation photon density $x_{ph} = 0.0016–0.079$ ph/Cu. The cross-correlation profile corresponding to the time resolution is also shown by the yellow line in Fig. 2a. Photoirradiation decreases the reflectivity due to photocarrier generation. This bleaching signal reflects the dynamics of both the mid-gap-absorption and the Drude components[16,17]. Figures 2b–d show typical time characteristics of $\Delta R(t)/R$ for weak ($x_{ph} = 0.0049$ ph/Cu), medium ($x_{ph} = 0.024$ ph/Cu) and strong ($x_{ph} = 0.079$ ph/Cu) excitations, respectively. In the weak-excitation case, $\Delta R(t)/R$ initially decreases within the time resolution (step A), further decreases with a rise time of ~20 fs (step B), and is followed by a slow decay of ~350 fs (step C). In $Nd_{2-x}Ce_xCuO_4$ with $x \leq 0.025$, a metallic behavior is not observed and the system is semiconducting[3,4]. Therefore, step B is related to the magnetic-polaron formations.

In the strong-excitation case (Fig. 2d), $\Delta R(t)/R$ reaches a minimum just after photoirradiation (step A). Subsequently, part of $\Delta R(t)/R$ decays within ~30 fs (step B') and the residual part decays within ~350 fs similar to the weak excitation case (step C). The former ultrafast-decay component gradually increases with increasing $x_{ph}$ (Fig. 2a) and is attributed to the Drude component, which is generated from the overlapping wavefunctions of electron (hole) carriers. In the negative time region ($t < 0$), the $\Delta R(t)/R$ signal appears at $t = -100$ fs and grows as $t$ approaches 0. This can be ascribed to the free induction decay (FID) or the reduction of probe-pulse-induced polarization by the pump pulse[23] (Supplementary Information). This component is not related to the photocarrier dynamics discussed here.

**Analyses of the time evolutions of the reflectivity changes.**

To analyze the time evolution of $\Delta R(t)/R$, the following formula is adopted in the weak excitation case.

$$\frac{\Delta R(t)}{R} = -\left\{A_1 + A_2\left[1 - \exp\left(-\frac{t}{\tau_r}\right)\right]\right\}\exp\left(-\frac{t}{\tau_1}\right) \tag{1}$$

The first and second terms show bleaching due to the initial photocarrier generation (step A) and the formation of the mid-gap states with time constant $\tau_r$ (step B), respectively. $\tau_1$ is the decay time of the mid-gap states (step C). $\Delta R(t)/R$ for $x_{\text{ph}} = 0.0049$ ph/Cu can be well reproduced by equation (1) with $\tau_r = 18$ fs as shown by the solid line in Fig. 2b, except for the FID components. The maximum energy of the optical phonons in Nd$_2$CuO$_4$ is 63 meV[24]. Consequently, $\tau_r = 18$ fs, which corresponds to 220 meV, cannot be explained by the lattice dynamics. Considering the magnetic-polaron formation in Nd$_{2-x}$Ce$_x$CuO$_4$, it is reasonable to ascribe $\tau_r = 18$ fs to the spin-relaxation time. Spin relaxations cause a carrier to become more mobile by reducing the surrounding antiferromagnetic spin couplings, which increases the bleaching signal.

To analyze $\Delta R(t)/R$ in the medium and strong excitation case (Figs. 2c and d, respectively), we add the term $-A_3\exp(-t/\tau_2)$ to equation (1) to express the Drude component with decay time $\tau_2$ (step B'). Upon considering the local nature of the spin relaxation processes, the analyses use the same values of $\tau_r$ and $A_2/A_1$ as those obtained for the weak excitation case. The time characteristic of $\Delta R(t)/R$ for $x_{\text{ph}} = 0.024$ and 0.079 ph/Cu are well reproduced as shown by black lines in Figs. 2c and d. The Drude and mid-gap-absorption components are shown by red and blue lines, respectively, in the same figures.

$\Delta R(t)/R$ for various $x_{\text{ph}}$ values can also be reproduced by the same formula (Supplementary Information). The $\Delta R(t)/R$ signals reflecting the Drude components thus derived are shown in Fig. 2e. The values of the parameters $A_3$ and $\tau_2$, which reflect the number of free carriers and their decay times, are plotted as a function of $x_{\text{ph}}$ in Figs.

2f and g, respectively. The number of free carriers ($A_3$) shows a clear threshold behavior, which is characteristic of two-dimensional Mott insulators[17]. As the carriers contributing to the Drude component increase, their decay time $\tau_2$ decreases. These behaviors suggest that an Auger recombination[25] illustrated in the inset of Fig. 2g plays a dominant role on the ultrafast decay of the metallic state. Detailed analyses of the long-time data ($t > 100$ fs) reveal that when the carrier number becomes small, the residual electron and hole carriers recombine with $\tau_1 \sim 350$ fs (Supplementary Information), which is thought to occur via the theoretically predicted multi-magnon emission[26,27].

**Probe-energy dependence on the time evolutions of the reflectivity changes.**

To derive the detailed information about the magnon side band originating from the charge-spin coupling, we next measured the probe-energy dependence on $\frac{\Delta R(t)}{R}$ (see Methods). Figure 3a shows the time evolution of $\frac{\Delta R(t)}{R}$ for the four probe energies (1.88 eV, 1.94 eV, 2.03 eV, and 2.10 eV). The excitation photon density is $x_{\text{ph}} = 0.008$ ph/Cu. This corresponds to the weak excitation case in which the Drude component can be neglected. All the time profiles for the four probe energies contain the high-frequency oscillatory structures. We extracted those oscillatory components $\frac{\Delta R_{\text{osc}}(t)}{R}$ from $\frac{\Delta R(t)}{R}$ using a Fourier filter, which are plotted in Fig. 3b. The Fourier power spectra of $\frac{\Delta R_{\text{osc}}(t)}{R}$ are also shown in Fig. 3c. The oscillation frequencies ($\hbar\Omega_n$) depend on the probe energy ($\hbar\omega_r$), which is an unusual behavior. Figure 3d plots $\hbar\Omega_n$ as a function of $\hbar\omega_r$ together with the $\sigma$ spectrum. $\hbar\Omega_n$ increases from the Mott-gap transition peak up to 2700 cm$^{-1}$, following the relation $\hbar\Omega_n = \hbar\omega_r - 1.74$ eV, which is shown by the broad gray line in Fig. 3d. 1.74 eV is close to the Mott-gap transition energy $\hbar\omega_{\text{CT}}$. This suggests that the oscillations may be related to the emission of two magnons. The origin of these oscillations is discussed in detail in the next section.

**Discussion**

Here, we discuss the origin of the high-frequency coherent oscillations observed on the reflectivity changes after the photoirradiation. In cuprate Mott insulators, two-magnon emission signatures are observed in the Raman scatterings[28]. It is therefore natural to consider that the two-magnon sideband is included on the higher-energy side of the Mott-gap transition peak. Figures 3e and f illustrate this sideband and the magnon dispersion[29], respectively. In the emission of two magnons with positive and negative momentums, the total spin is conserved. Thus, the two-magnon frequency $\hbar\Omega_n$ is connected to the probe energy $\hbar\omega_r$ via $\hbar\omega_r = \hbar\omega_{\text{CT}} + \hbar\Omega_n$ as observed in the experiments.

To explain the generation of the coherent oscillations, we consider the photoexcited states $|N, n\rangle$, where $N$ (=1 or 2) denotes the number of doublon-holon pairs and $n$ denotes the two-magnon excited state with the frequency $\hbar\Omega_n$. A broadband pump pulse generates $|1, n\rangle$ with various frequencies $\hbar\Omega_n$ and a probe pulse produces additional transitions (e.g., $|1, 0\rangle \rightarrow |2, n\rangle$ and $|1, n\rangle \rightarrow |2, n\rangle$), which interfere with each other. This results in a coherent oscillation with the frequency $\hbar\Omega_n$, which corresponds to the energy difference between the initial photoexcited states $|1, 0\rangle$ and $|1, n\rangle$. The coherent oscillation should be detected at the probe energy of $\hbar\omega_r = \hbar\omega_{\text{CT}} + \hbar\Omega_n$, which is the energy for $|1, n\rangle$. The results and the theoretical simulation of the oscillations are reported in the Supplementary Information.

Finally, we summarize the photocarrier relaxation processes in $Nd_2CuO_4$, which is schematically shown in Fig. 4. In the weak excitation case, photocarriers form magnetic polarons in ~18 fs (a → b → d in Fig. 4). In the strong excitation case, a number of carriers are generated just after photoirradiation (a → c in Fig. 4). The system becomes metallic because the wavefunctions of the photocarriers overlap and the metallic state decays in ~10–40 fs (c → d in Fig. 4), depending on the initial photocarrier densities. An Auger recombination is a plausible decay process. The residual small numbers of photocarriers form magnetic polarons. The coherent oscillations observed in the pump-probe responses are attributed to quantum interferences between charge-spin coupled excited states. Thus, sub-10 fs optical spectroscopy is a powerful tool to capture the spin dynamics as well as charge dynamics under the presence of strong charge-spin coupling.

**Methods**

**Sample preparations.**

$Nd_2CuO_4$ single crystals were grown in air by a modified self-flux method[30]. A mixture of 200 grams of $Nd_2O_3$ and CuO at a molar ratio of Nd:Cu = 28:72 was filled in a Pt crucible. The crucible was heated at 1300 °C for 2 hours in a furnace to melt the mixture completely. Then the tip of a Pt wire was dipped in the melt, and the temperature was lowered at a speed of 3 °C/h to 1220 °C. The crystals grown around the tip were raised by pulling up the wire. The melt around the crystals was dripped, forming plate-like crystals with shiny flat surfaces.

**Steady-state optical spectroscopy measurements.**

The optical reflectivity ($R$) spectrum in a single crystal of $Nd_2CuO_4$ was measured using a specially designed spectrometer with a 25-cm-grating monochromator for 0.5-5.0 eV and a Fourier-transform infrared spectrometer for 0.15-1.2 eV. Both were equipped with an optical microscope. The optical conductivity ($\sigma$) spectrum was calculated from the $R$ spectrum by the Kramers-Kronig transformation.

**Femtosecond pump-probe reflection measurements.**

The light source for the pump-probe (PP) measurements was a Ti:sapphire regenerative amplifier with a central wavelength of 800 nm, pulse width of 130 fs, pulse fluence of 0.8 mJ, and repetition rate of 1 kHz. To generate an ultrashort pulse, we self-produced a non-collinear optical parametric amplifier[31,32]. We divided the output of the regenerative amplifier into two beams. One was converted to second harmonic (SH) light using a $LiB_3O_5$ crystal. The other was focused on a sapphire crystal, from which a broadband white-light pulse was generated. This white-light pulse was used as a seed and amplified by the SH light in a $\beta$-$BaB_2O_4$ crystal. The pulse width at the sample position was compressed by a pair of chirp mirrors and quartz plates inserted between the sapphire crystal and the sample. Finally, an ultrashort pulse with a temporal width of 7 fs and a spectrum width of ~0.5 eV (Fig. 1b) was obtained.

The setup of the PP reflectivity measurements is illustrated in the inset of Fig. 1b. The 7-fs pulse was divided into two pulses, which were used as pump and probe pulses. The polarizations of the pump and probe pulses were the same and parallel to either the *a* or *b* axis. A variable delay stage controlled the delay time of the probe pulse relative to the pump pulse. To suppress the effects of the intensity fluctuations of the probe pulse on the signals, we normalized the reflection intensity of each probe pulse by the intensity of the pulse before the incidence, which was measured by extracting part of the probe pulse using a semitransparent mirror. The cross-correlation profile of the pump and probe pulses, which is shown by the yellow line with the yellow shading in Fig. 2a. The full width at the half maximum was 10.0 fs, corresponding to the time resolution. In the measurements of the probe-energy dependence of the reflectivity changes, we selected part of the reflection light by inserting a band-pass filter (bandwidth of 10 nm and central wavelengths of 588 nm, 610 nm, 636 nm, or 656 nm) in front of the detector (a photodiode).

The excitation photon density per Cu site, $x_{\text{ph}}$, was evaluated from the formula $x_{\text{ph}} = I_{\text{p}}(1 - R_{\text{p}})(1 - 1/e)/l_{\text{p}}$, in which $I_{\text{p}}$, $R_{\text{p}}$, and $l_{\text{p}}$ are the photon density per unit area, reflectivity, and penetration depth of the pump pulse, respectively. $l_{\text{p}}$ was evaluated from the absorption-coefficient spectrum, which was obtained by the Kramers-Kronig analysis of the *R* spectrum.

**Data availability.**

The data that support the plots within this paper and other findings of this study are available from the corresponding author upon reasonable request.

**Acknowledgments**

We thank Mr. Y. Miyata, Mr. N. Osawa and Mr. M. Inoue for their collaborations in the early stage of this study. This work was partly supported by Grants-in-Aid for Scientific Research from the Japan Society for the Promotion of Science (JSPS) (Project Numbers: 15H06130 and 16K17721) and by CREST (Grant Number: JPMJCR1661), Japan Science and Technology Agency. T.T. and T.Morimoto were supported by the JSPS through the Program for Leading Graduate Schools (MERIT).

**Author contributions**

T.Miyamoto., Y.M., T.Morimoto, N.S., Y.W., and S.A. built the experimental apparatus. T.Miyamoto., Y.M., T.T., N.S., and H.Y carried out the transient optical measurements. T.I, K.O. and A.S. prepared the samples. S.I. performed theoretical calculations. H.O. coordinated the study. All of the authors discussed the results and contributed to writing the paper.

**Additional information**

Supplementary information is available in the online version of the paper. Reprints and permissions information is available online at www.nature.com/reprints.

Correspondence and requests for materials should be addressed to H.O.

**Competing financial interests**

The authors declare no competing financial interests.

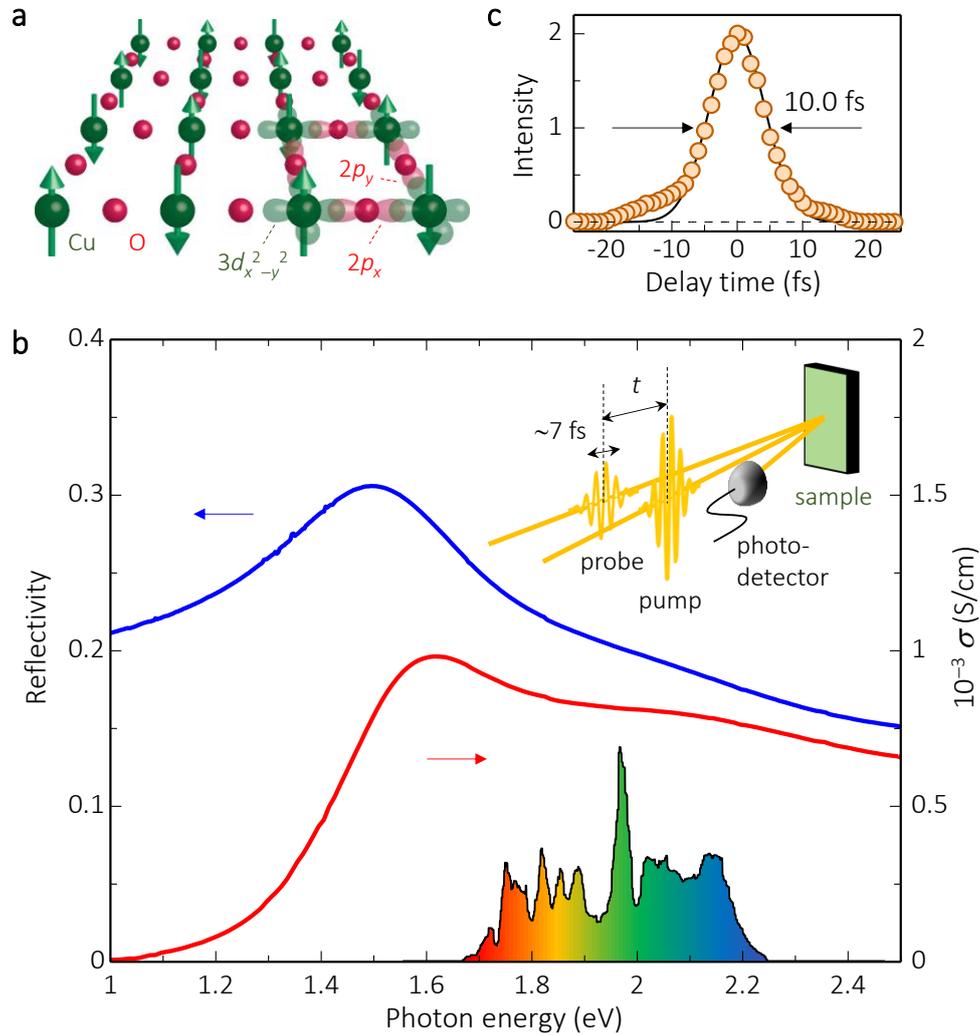

**Figure 1 | Optical spectra and pump-probe reflection measurements on CuO$_2$ planes in Nd$_2$CuO$_4$. a,** The two-dimensional electronic state and the spin arrangement (arrows) in a CuO$_2$ plane (the *ab* plane). **b,** The reflectivity (*R*) and optical conductivity (*σ*) spectra measured on the *ab* plane. Electric fields of lights are polarized parallel to *a* or *b* axis. The black line in the lower part shows a spectrum of a 7-fs pulse used in the pump-probe (PP) experiments. The inset shows a schematic of PP experiments. **c,** A cross-correlation profile of pump and probe pulses corresponding to the time resolution.

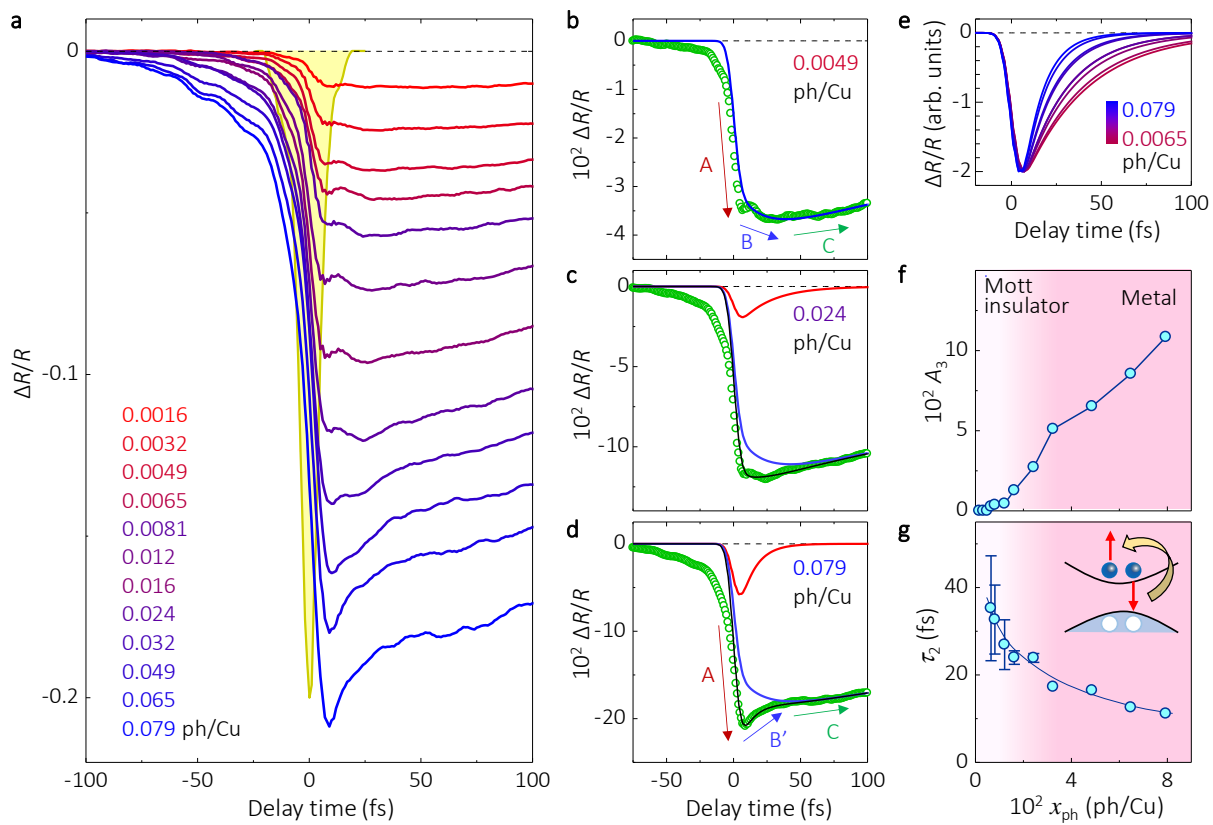

**Figure 2 | Results of the pump-probe reflection measurements in Nd$_2$CuO$_4$. a,** Time evolutions of reflectivity changes $\Delta R/R$ for $x_\mathrm{ph} = 0.0016 - 0.079$ ph/Cu. A yellow line reshows the cross-correlation profile between pump and probe pulses. **b-d,** Time evolutions of $\Delta R/R$ for (**b**) $x_\mathrm{ph} = 0.0049$ ph/Cu, (**c**) $x_\mathrm{ph} = 0.024$ ph/Cu, and (**d**) $x_\mathrm{ph} = 0.079$ ph/Cu with fitting curves. Blue and red lines reflect the mid-gap-absorption components and the Drude components, respectively. Thin black lines show the sum of two components. **e,** Time evolutions of reflectivity changes $\Delta R/R$ reflecting the Drude components derived from the fitting analyses (red lines in Fig. **c,d**). **e,f,** $x_\mathrm{ph}$-dependences of (**e**) amplitudes and (**f**) decay time of the reflectivity changes $\Delta R/R$ reflecting the Drude components. The inset shows a schematic illustration of the Auger recombination.

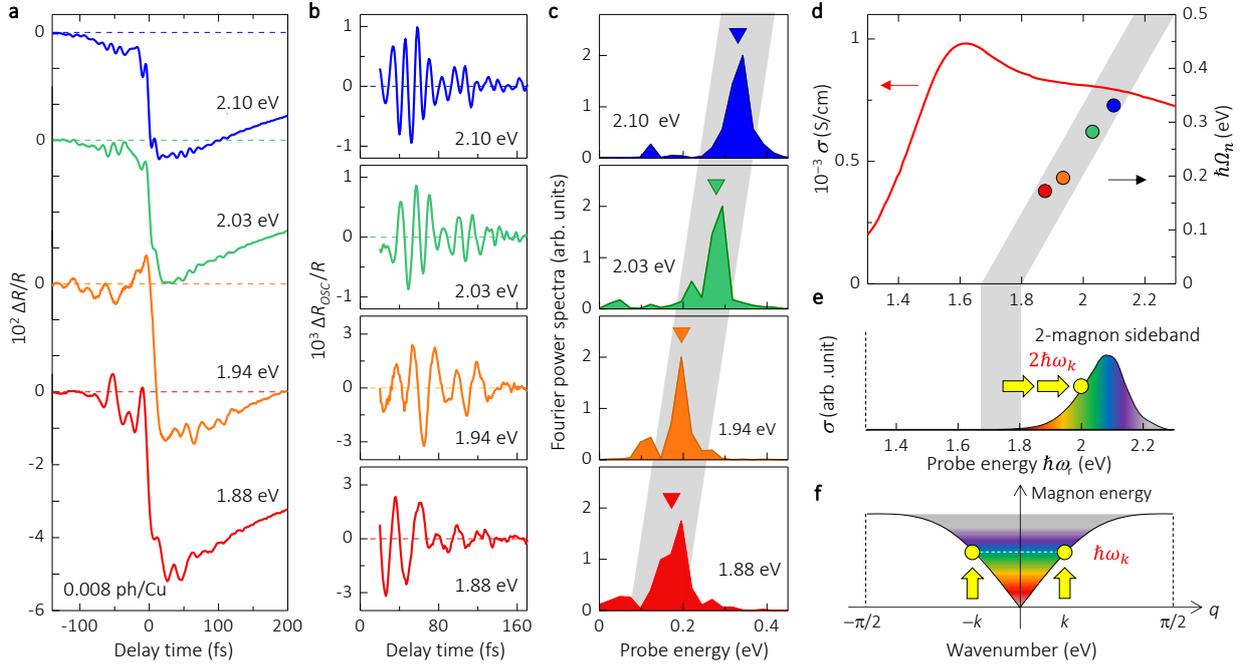

**Figure 3 | Probe energy dependence of reflectivity changes in Nd$_2$CuO$_4$. a,** Time evolutions of $\Delta R/R$ for $x_{\mathrm{ph}} = 0.008$ ph/Cu at various probe energies $(\hbar\omega_{\mathrm{r}})$. **b,** Oscillatory components $\Delta R_{\mathrm{OSC}}/R$ extracted from $\Delta R/R$. **c,** Fourier power spectra of $\Delta R_{\mathrm{OSC}}/R$. Triangles indicate the oscillation frequencies $\hbar\Omega_n$. **d,** Probe energy dependence of $\hbar\Omega_n$. The solid line shows the $\sigma$ spectrum. The broad gray line is drawn from 1.73 eV ($\sim\hbar\omega_{\mathrm{CT}}$) with the slope of 1 (see the text). **e,** A two-magnon sideband is schematically shown as the shaded area. **f,** Schematics of the magnon dispersion.

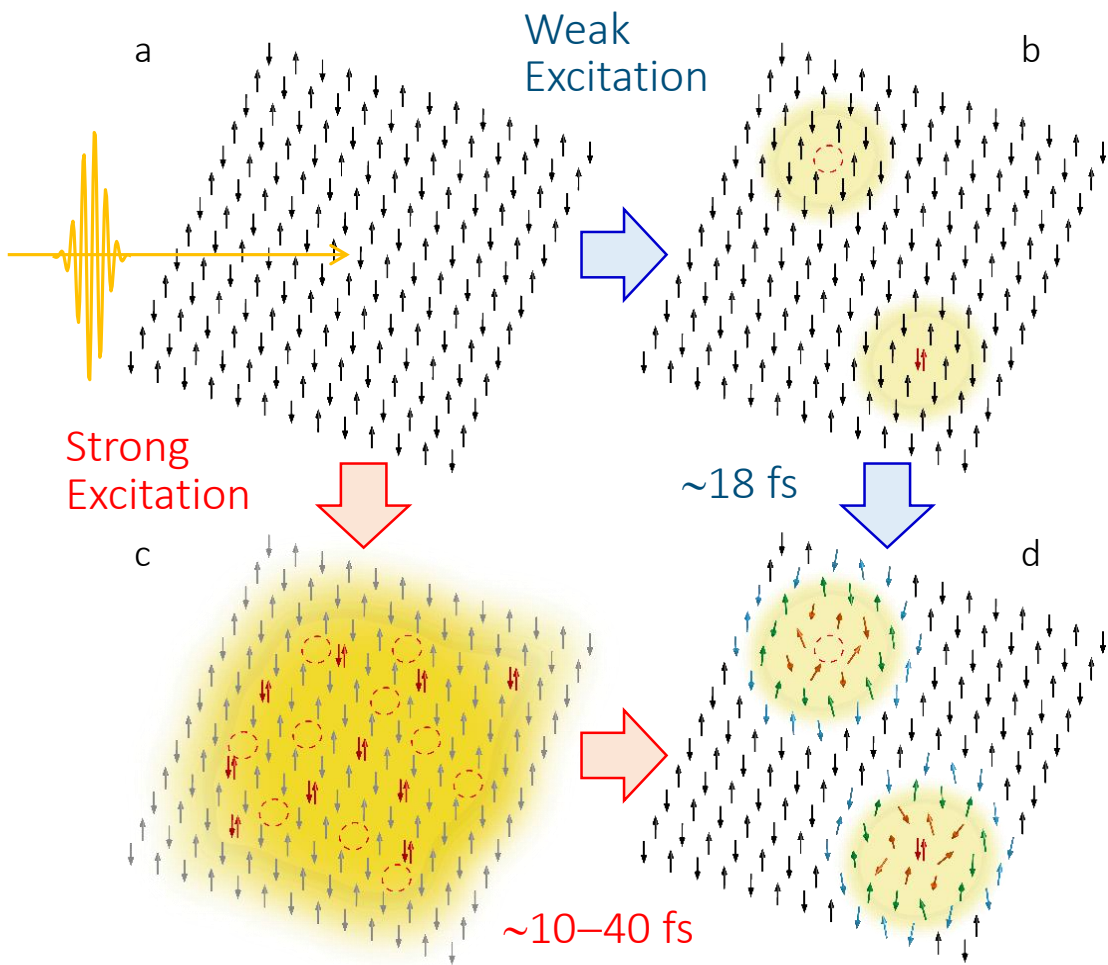

**Figure 4 | A schematic of the ultrafast photoresponses in Nd$_2$CuO$_4$. a,** The Mott insulator state. **b,c,** Generation of (**b**) low density and (**c**) high density of doublons and holons just after the photoirradiation. **d,** Formations of magnetic polarons in the relaxation process.

**Supplementary Material:**

**Probing ultrafast spin-relaxation and precession dynamics in a cuprate Mott insulator with 7-fs optical pulses**

T. Miyamoto[1*], Y. Matsui[1], T. Terashige[2], T. Morimoto[1], N. Sono[1], H. Yada[1], S. Ishihara[3], Y. Watanabe[4], S. Adachi[4], T. Ito[5], K. Oka[5], A. Sawa[5], H. Okamoto[1,2*]

[1]*Department of Advanced Materials Science, University of Tokyo, Chiba 277-8561, Japan.*

[2]*AIST-UTokyo Advanced Operando-Measurement Technology Open Innovation Laboratory (OPERANDO-OIL), National Institute of Advanced Industrial Science and Technology (AIST), Chiba 277-8568, Japan*

[3]*Department of Physics, Tohoku University, Sendai 980-8578, Japan.*

[4]*Department of Chemistry, Kyoto University, Kitashirakawa Oiwake-cho, Sakyo-ku Kyoto, 606-8502, Japan.*

[5]*National Institute of Advanced Industrial Science and Technology, Tsukuba, Ibaraki 305-8565, Japan*

*Corresponding author: T. Miyamoto (miyamoto@k.u-tokyo.ac.jp), H. Okamoto (okamotoh@k.u-tokyo.ac.jp)

**Content**

**S1. Free induction decay signals and oscillatory components**

**S2. Fitting analyses of the time evolution of reflectivity changes**

**S3. Time evolutions of the photoinduced reflectivity changes in the long-time domain**

**S4. Theoretical analyses of coherent oscillations on reflectivity changes**

## S1. Free induction decay signals and oscillatory components

The photoinduced reflectivity changes $\Delta R/R$ for $t<0$ (Figs. 2a and 3a) are attributed to the free induction decay (FID) signals, which are explained as follows. A probe pulse generates a polarization, which emits an electromagnetic wave or equivalent light with the polarization frequency. This is observed as the reflected probe light. A pump pulse reaching the sample after the probe pulse modulates the probe-pulse-induced polarization, resulting in the modulation of the reflected probe light. With the decrease of $|t|$ at $t<0$, such a modulation should be enhanced, which is consistent with the observations.

In the probe-energy dependence of $\Delta R/R$ (Fig. 3a), the oscillatory components appear not only at $t>0$ but also at $t<0$. By applying a Fourier filter to the time evolutions of $\Delta R/R$, we extracted the oscillatory components ($\Delta R_{OSC}/R$) at $t<0$ and $t>0$, which are shown in Figs. S1a and b, respectively. The oscillations in the FID signals were previously reported in organic materials and ascribed to coherent phonons coupled with the photoexcited state[1,2]. In $Nd_2CuO_4$, the spin system is strongly coupled with the photoexcited state. Consequently, the oscillatory component in the FID signals can also be assigned to the spin excitations or magnons. The Fourier power spectra of $\Delta R_{OSC}/R$ are shown in Fig. S1c. The peak energy of each oscillatory component at $t<0$ (orange lines) is almost equal to that at the same probe photon energy (blue lines). Therefore, the oscillatory components at $t<0$ and $t>0$ have the same origin.

## S2. Fitting analyses of the time evolution of reflectivity changes

To analyze the time evolution of $\Delta R(t)/R$, we adopted the fitting function equation (S1), which is given as

$$\frac{\Delta R(t)}{R} = -\left\{A_1 + A_2\left[1 - \exp\left(-\frac{t}{\tau_r}\right)\right]\right\}\exp\left(-\frac{t}{\tau_1}\right) - A_3\exp\left(-\frac{t}{\tau_2}\right) \quad (S1)$$

The first term on the right-hand side represents the mid-gap-absorption component ascribed to magnetic polarons. The second term, $-A_3\exp\left(-\frac{t}{\tau_2}\right)$, shows the formation and decay of the Drude component, which is neglected in equation (1). The main text describes the physical meanings of the parameters. In the fitting procedures, each term in equation (S1) or equation (1) is convolved with the Gaussian profile corresponding to the time resolution (10 fs). From the values of $A_3$ and $\tau_2$, the excitation photon density dependence of the magnitude and the decay time of the Drude component can be discussed. For a weak excitation density, $x_{\text{ph}}$ = 0.0016, 0.0032, and 0.0049 ph/Cu, the Drude component can be neglected.

Figure S2a-l replots all of the time evolutions of the reflectivity changes displayed in Fig. 2a by green open circles, which are well reproduced by the fitting curves (thin black lines) except for $t < 0$. The time evolutions of the first and second terms are represented in Fig. S2m-x by blue and red lines, respectively. The values of parameters $A_3$ and $\tau_2$ are shown in Figs. 2f and g, respectively.

**S3. Time evolutions of the photoinduced reflectivity changes in the long-time domain**

In Figs. 2 and S2, we analyzed the time evolutions of the photoinduced reflectivity changes $\Delta R/R$ from $t = -100$ fs to $t = 200$ fs to obtain the fitting parameters. To evaluate more precisely the recombination time of carriers (magnetic polarons), $\tau_1$, we measured $\Delta R/R$ in the long-time region up to 500 fs. Figure S3 shows the results for both the weak excitation ($x_{\text{ph}} = 0.0049$ ph/Cu) and strong excitation ($x_{\text{ph}} = 0.079$ ph/Cu). The data in the short-time region are the same as those presented in Figs.

S2c and l. In $\Delta R/R$ at $t > 100$ fs, a component with a long decay time exists. This component may be attributed to heating of the system, which originates from the recombination of carriers[3,4]. To include this thermal-effect component in the analyses, the third term is added to equation (S1).

$$\frac{\Delta R(t)}{R} = -\left\{A_1 + A_2\left[1 - \exp\left(-\frac{t}{\tau_r}\right)\right]\right\}\exp\left(-\frac{t}{\tau_1}\right)$$
$$-A_3\exp\left(-\frac{t}{\tau_2}\right) - A_4\left[1 - \exp\left(-\frac{t}{\tau_1}\right)\right] \quad (S2)$$

It is reasonable to consider that the rise time of the thermal-effect component is the same as the decay time of carriers $\tau_1$[4]. In the fitting procedures, we used the same parameter values of $A_2/A_1$, $\tau_r$, and $\tau_2$ as those used in Figs. S2c(o) and l(x). The black lines in Fig. S3 show the fitting curves. The decay dynamics in the long-time region can be well reproduced with $\tau_1 = 350$ fs.

## S4. Theoretical analyses of coherent oscillations on reflectivity changes

To explain the generation of the coherent oscillations in the pump-probe responses, we considered a simple semi-phenomenological model given by

$$H_0 = \hbar\omega_{CT}X^\dagger X + \sum_k \hbar\omega_k b_k^\dagger b_k, \quad (S3)$$

where $X^\dagger(X)$ is the creation (annihilation) operators for a doublon-holon pair with an energy of $\hbar\omega_{CT}$ and $b_k^\dagger(b_k)$ is a magnon with the momentum $k$ and the energy of $\hbar\omega_k$. The electronic and spin states are represented by $|N, n\rangle$, where $N$ (=1 or 2) denotes the number of doublon-holon pairs and $n$ denotes a two-magnon excited state with a frequency of $2\hbar\omega_n$, respectively. For simplicity, a photon with the frequency $\omega$ is assumed to excite specifically a doublon-holon pair and two-magnons with the energy of $2\hbar\omega_k$, which is expressed by the following Hamiltonian

$$H' = A_{\omega_{\text{CT}}} X^\dagger + g \sum_k A_{\omega_{\text{CT}}+2\omega_k} X^\dagger b_k^\dagger b_{-k}^\dagger + H.c. \tag{S4}$$

$A_\omega$ is the component of the vector potential with frequency $\omega$, and $g$ is the coupling constant. A pump pulse introduced at $t = 0$ generates a photoexcited state $|\Psi(0)\rangle$, which is represented by a linear combination of the initial state $|0\rangle$, one doublon-holon pair state $|1,0\rangle$, and the two-magnon sideband state $|1,n\rangle$. The electronic current coupled with the vector potential is identified from $H'$ in equation (S4). The expectation of the electronic current $j$ at $t$, $\langle j\rangle(t)$, is calculated as $\langle j\rangle(t) = \langle\Psi(t)|j|\Psi(t)\rangle$ using the time-dependent wave function $|\Psi(t)\rangle = e^{-iH_0 t}|\Psi(0)\rangle$.

Here we demonstrate that the coherent oscillation appears in the simplified model from equation (S3), where the continuous magnon spectrum is replaced by two discrete energy levels labeled as $K$ and $Q$. The corresponding two-magnon energies are given by $\hbar\Omega_K \,(= 2\hbar\omega_K)$ and $\hbar\Omega_Q (= 2\hbar\omega_Q > \hbar\Omega_K)$, respectively. The energy levels of the excited states considered in this model are shown in Fig. S4a. The current expectation value is approximately given as $\langle j\rangle(t) \sim \int_0^t dt' F(t')$ with

$$\begin{aligned}
F(t') &= w_1 \sin\omega_{\text{CT}}(t-t')\, A'_{\omega_{\text{CT}}}(t')\\
&\quad + w_2 \sin[(\omega_{\text{CT}}+\Omega_K)(t-t')]\, A'_{\omega_{\text{CT}}+\Omega_K}(t')\\
&\quad + w_3 \sin[(\omega_{\text{CT}}+\Omega_Q)(t-t')]\, A'_{\omega_{\text{CT}}+\Omega_Q}(t')\\
&\quad + w_4\{\sin[\omega_{\text{CT}}t-(\omega_{\text{CT}}+\Omega_K)t']\, A'_{\omega_{\text{CT}}+\Omega_K}(t')\\
&\qquad + \sin[(\omega_{\text{CT}}+\Omega_K)t-\omega_{\text{CT}}t']\, A'_{\omega_{\text{CT}}}(t')\}\\
&\quad + w_5\{\sin[\omega_{\text{CT}}t-(\omega_{\text{CT}}+\Omega_Q)t']\, A'_{\omega_{\text{CT}}+\Omega_Q}(t')\\
&\qquad + \sin[(\omega_{\text{CT}}+\Omega_Q)t-\omega_{\text{CT}}t']\, A'_{\omega_{\text{CT}}}(t')\}\\
&\quad + w_6\{\sin[(\omega_{\text{CT}}+\Omega_K)t-(\omega_{\text{CT}}+\Omega_Q)t']\, A'_{\omega_{\text{CT}}+\Omega_Q}(t')\\
&\qquad + \sin[(\omega_{\text{CT}}+\Omega_Q)t-(\omega_{\text{CT}}+\Omega_K)t']\, A'_{\omega_{\text{CT}}+\Omega_K}(t')\},
\end{aligned} \tag{S5}$$

where $w_l$ ($l = 1 - 6$) is a constant and $A'_\omega$ is the vector potential for the probe photon. The 4th term (5th term) represents the interference between the processes of $|1,0\rangle \to |2,K\rangle$ and $|1,K\rangle \to |2,K\rangle$ ($|1,0\rangle \to |2,Q\rangle$ and $|1,Q\rangle \to |2,Q\rangle$), which is schematically shown in Fig. S4b. When $t \sim t'$, these terms show a coherent oscillation with frequency $\Omega_K$ ($\Omega_Q$). The last two terms represent the interferences between the processes of $|1,K\rangle \to |2,K,Q\rangle$ and $|1,Q\rangle \to |2,K,Q\rangle$. In an actual situation where two-magnon states are continuous, these terms do not give coherent oscillations.

To show a coherent oscillation due to these interferences, the probe vector potential is set to be a damped oscillator given as $A'_\omega(t) = f_\omega e^{-(t-t_0)^2/\tau^2} \cos\omega(t-t_0)$ with $f_\omega = e^{-(\omega-\omega_0)^2/\delta\omega^2}$. Figure S4c,d shows parts of $\langle j\rangle(t)$ where the 4th and 5th terms in equation (S5) (i.e., the interference terms) are adopted in the calculations. Fourier transforms of Figs. S4c and d are shown in Figs. S4e and f, respectively. The frequencies of the probe photons are tuned at $\omega_0 = \omega_{\text{CT}} + \Omega_K$ and $\omega_0 = \omega_{\text{CT}} + \Omega_Q$ in Figs. S4c and d, respectively. The parameter values are chosen as $\Omega_K = 1, \Omega_Q = 2, \omega_{\text{CT}} = 10, \tau = 0.5$, and $\delta\omega = 1$. The results show that the current spectra have a peak structure near $\Omega_K$ ($\Omega_Q$), when the probe photons are tuned around $\omega_{\text{CT}} + \Omega_K$ ($\omega_{\text{CT}} + \Omega_Q$), giving an explanation of the characteristic coherent oscillation experimentally observed.

**Supplementary References**

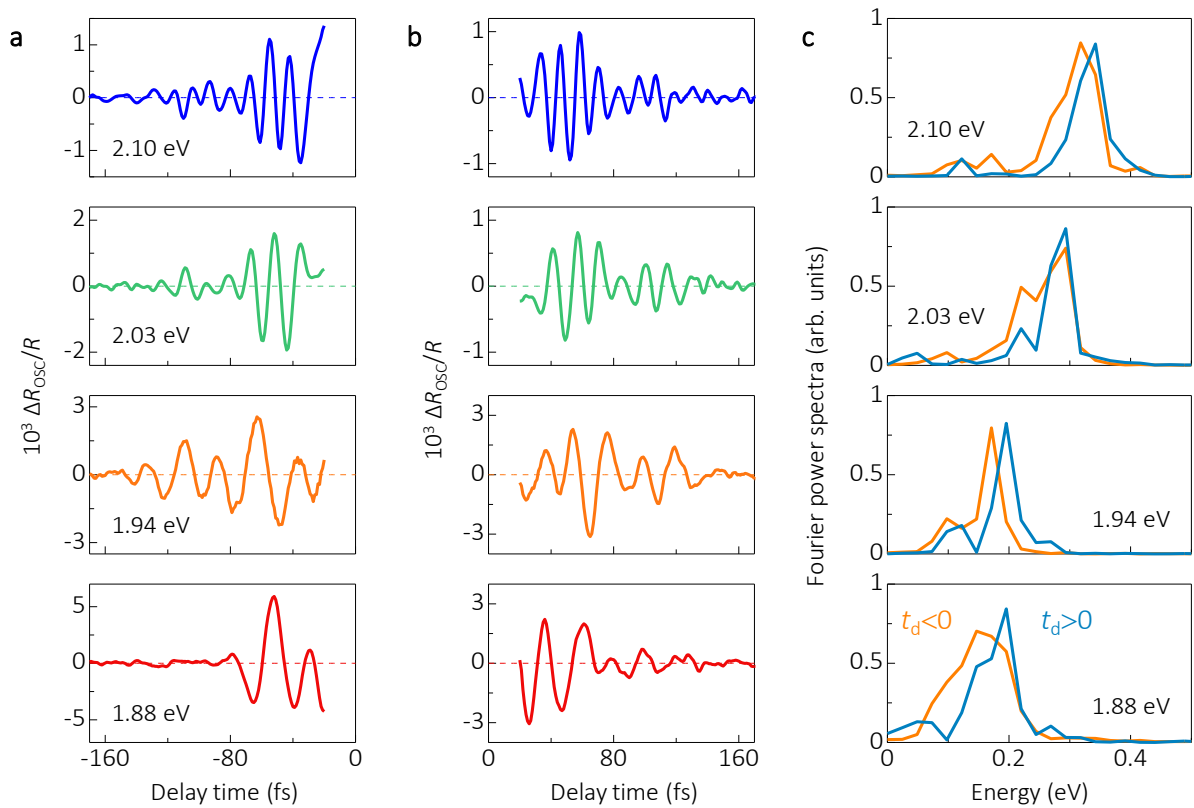

**Figure S1 | Probe energy dependence of the oscillations on $\Delta R/R$. a,b,** Time evolutions of the oscillatory components $\Delta R_{\mathrm{OSC}}/R$ (**a**) at $t<0$ and (**b**) at $t>0$ extracted from the data shown in Fig. 3a. The probe energies are shown in **a**. **c,** Fourier power spectra of $\Delta R_{\mathrm{OSC}}/R$ at $t<0$ shown in Fig. **a** (orange lines) and at $t>0$ shown in Fig. **b** (blue lines).

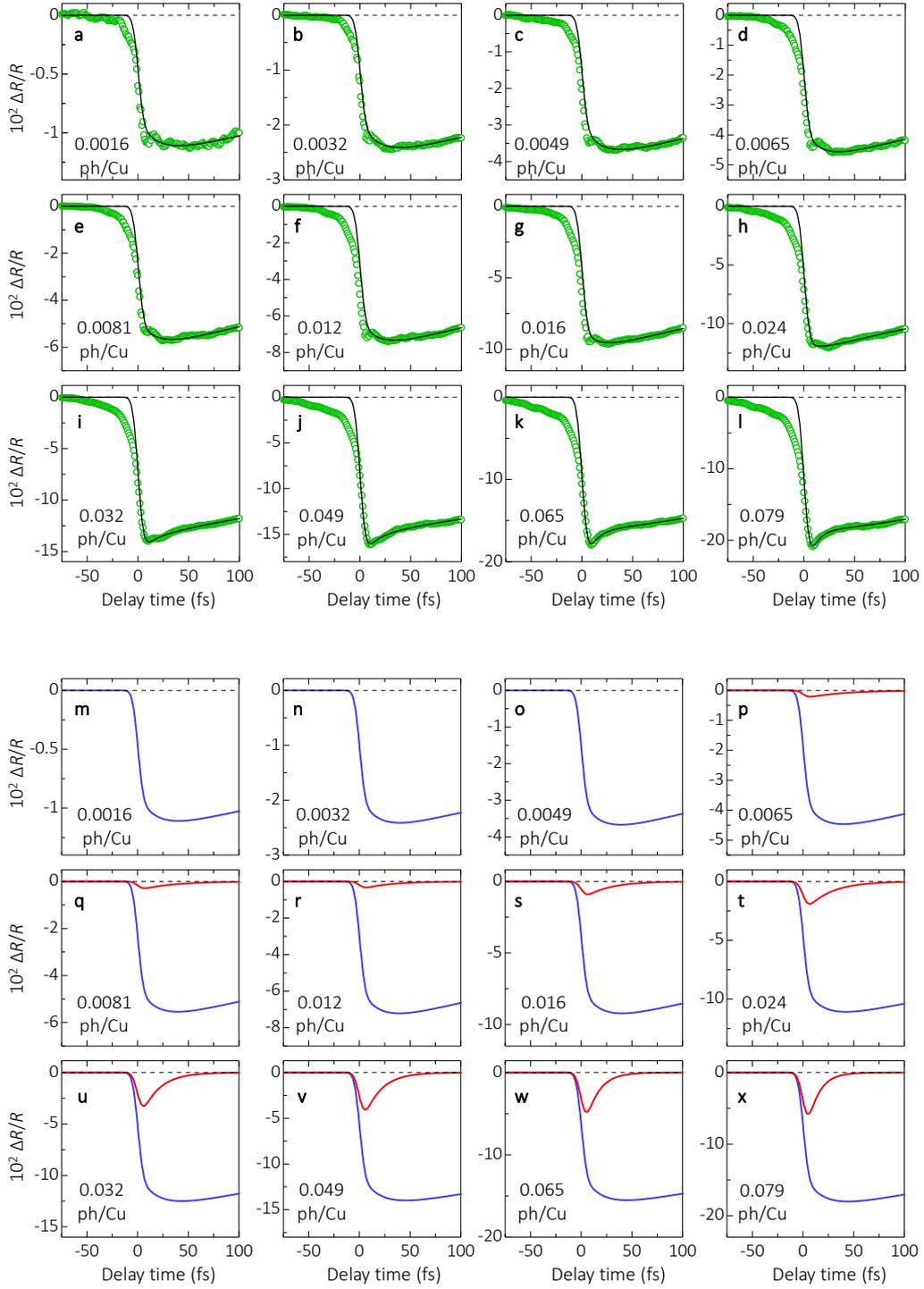

**Figure S2 | Fitting analyses of the time evolutions of $\Delta R/R$. a-l,** Open circles and black lines show time evolutions of $\Delta R/R$ and the fitting curves for $x_{ph}$ = 0.0016-0.079 ph/Cu. **m-x,** Blue and red lines represent the mid-gap-absorption components and the Drude components, respectively, which are obtained by the fitting analyses.

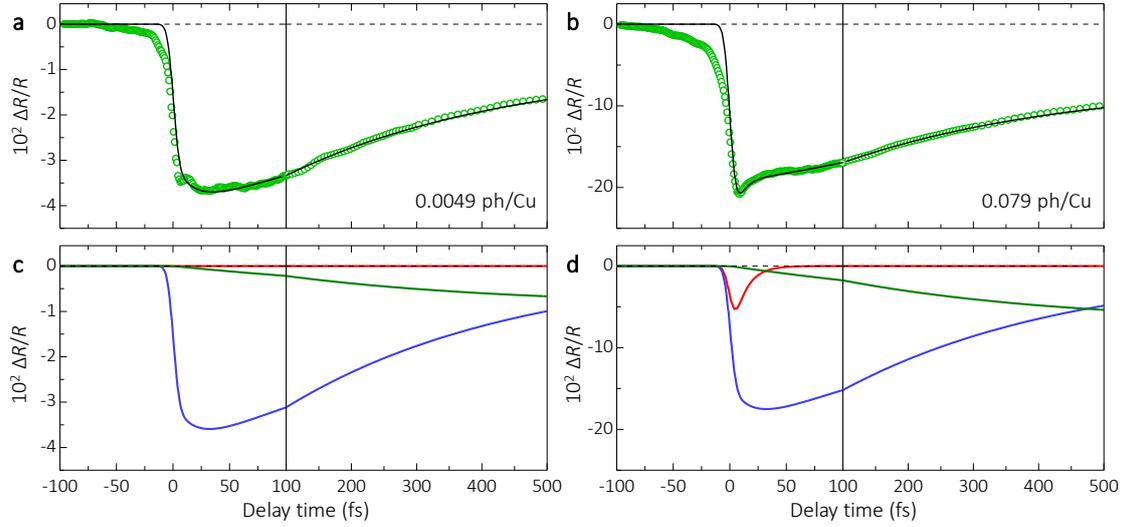

**Figure S3 | Analyses of Δ*R*/*R* in the long time region. a,b,** Green open circles and black lines show the time evolutions of $\Delta R/R$ and the fitting curves for the weak excitation ($x_\mathrm{ph} = 0.0049$ ph/Cu) in (**a**) and the strong excitation ($x_\mathrm{ph} = 0.079$ ph/Cu) in (**b**). **c,d,** Blue, red, and green lines represent the mid-gap-absorption component, the Drude component, and the thermal-effect component, respectively, obtained by the fitting analyses.

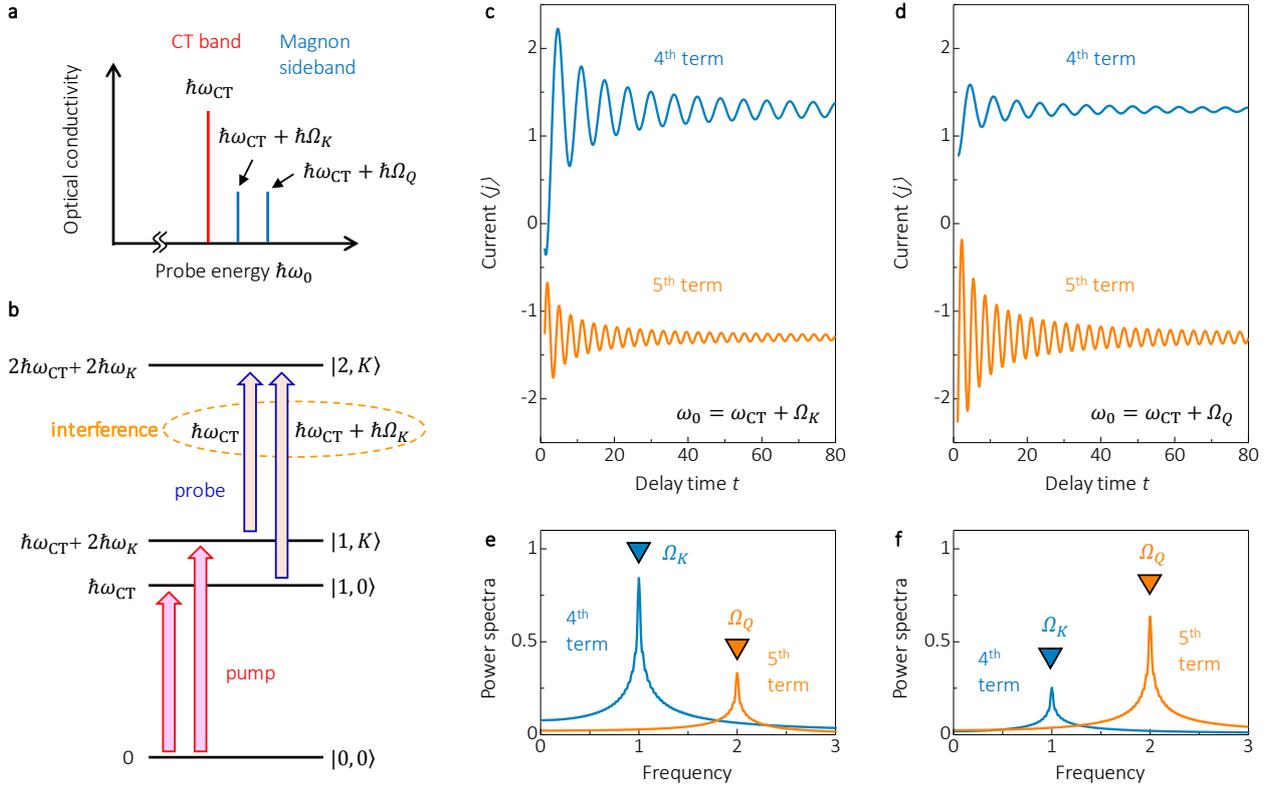

**Figure S4 | Time-dependent currents and Fourier power spectra calculated in a simplified system with two different frequencies of magnons. a,** The energy levels assumed in the model. **b,** A schematic of the interference between the processes of $|1,0\rangle \to |2,K\rangle$ and $|1,K\rangle \to |2,K\rangle$. **c,d,** Time dependent currents for the 4th term (blue lines) and the 5th term (orange lines) in equation (S5). They are vertically offset for clarity. The frequencies of the probe pulse are set at around $\omega_{CT} + \Omega_K$ in (**c**), and $\omega_{CT} + \Omega_Q$ in (**d**). **e,f,** Fourier power spectra of (**c**) in (**e**) and (**d**) in (**f**).

11